\documentclass[acmsmall]{acmart}

\AtBeginDocument{%
  }

\setcopyright{acmlicensed}
\copyrightyear{2024}
\acmYear{2024}
\acmDOI{XX.XX}

\acmConference[]{}{}{}
\acmISBN{XXX}
\usepackage{paralist}
\usepackage{wrapfig}
\usepackage{hyperref}
\usepackage{tabularx}
\usepackage{booktabs}
\usepackage{multirow}
\usepackage{array}
\usepackage{paralist}
\usepackage{enumitem}
\usepackage[utf8]{inputenc}
\usepackage{wrapfig}

\begin{document}

\title[Struggle Premium]{\textit{Struggle Premium}: How Human Effort and Imperfection Drive Perceived Value in the Age of AI}
\settopmatter{authorsperrow=4}
\author{Nazneen Sultana}
\affiliation{%
  \institution{Shahjalal University of Science and Technology}
  \department{Sociology}
  \city{Sylhet}
  \country{Bangladesh}}
\email{nazneensultana283@gmail.com}

\author{Mst Rafia Islam}
\affiliation{%
  \institution{Independent University, Bangladesh}
  \department{Department of Law}
  \city{Dhaka}
  \country{Bangladesh}}

\author{Md. Tanvir Hossain}
\affiliation{%
  \institution{Shahjalal University of Science and Technology}
  \department{Sociology}
  \city{Sylhet}
  \country{Bangladesh}}
\email{tanvir82@student.sust.edu}

\author{Azmine Toushik Wasi}
\orcid{0000-0001-9509-5804}
\affiliation{%
  \institution{Shahjalal University of Science and Technology}
  \department{Industrial Engineering}
  \city{Sylhet}
  \country{Bangladesh}}
\email{azmine32@student.sust.edu}

\renewcommand{\shortauthors}{Wasi et al.}
\setcopyright{none}

\begin{abstract}
As AI enters creative practice, audiences face growing uncertainty in judging authenticity and value. This study examines the \textit{Struggle Premium}, the added value attributed to perceived human effort, by analyzing how visible effort cues influence evaluations of human- and AI-generated creative works. We surveyed 70 university students, focusing on process videos, time documentation, written explanations, and imperfections. Process-oriented cues, especially videos and time spent, most strongly shaped authenticity and value judgments, while imperfections had limited impact. Participants showed a clear preference for human-made works, with 72.9\% willing to pay more. Notably, effort cues also improved perceptions of AI-generated content, suggesting that process transparency can partially bridge authenticity gaps. These findings extend the \textit{effort heuristic} to algorithmic creativity and inform the design of transparent human–AI creative systems.
\end{abstract}


%
%


\maketitle

\section{Introduction}

\label{sec:intro}
The rapid advancement of artificial intelligence (AI) in creative domains has challenged long-standing notions of artistic validity and value attribution. While generative AI systems are now capable of producing works that match or even surpass human technical proficiency, debates persist over their perceived authenticity, emotional depth, and market worth \cite{Bellaiche2023}. This tension has given rise to what we describe as the \textit{Struggle Premium}, a persistent societal preference for human-created art, even when AI alternatives achieve comparable aesthetic quality. The theoretical foundation for this phenomenon lies in the effort heuristic, a cognitive bias whereby observers infer quality and worth from perceived labor input rather than solely from the final output \cite{kruger2004effort}. Acting as a cognitive shortcut, this bias allows evaluators to rely on visible signs of effort instead of conducting in-depth assessments of the work itself. Experimental evidence supports this tendency: Bellaiche et al. \cite{Bellaiche2023} found that, even when controlling for aesthetic merit, participants systematically favored art produced by humans. Neuroimaging research adds further nuance, revealing that awareness of human authorship activates neural pathways associated with empathy and social cognition \cite{Hong2019-ym}. However, the rise of AI-assisted creativity disrupts these cues, blurring the lines between human and machine labor. While process transparency, such as: revealing AI training data, production steps, or collaborative roles can mitigate skepticism toward AI-generated works \cite{zerilli2022transparency}, emerging findings suggest such disclosures can also enhance appreciation for AI output \cite{Salas_Espasa2025-en}. Nevertheless, AI creations are still frequently perceived as lacking the emotional struggle, personal narrative, and authenticity that human effort imparts \cite{Chamberlain2018-am}.

Despite these insights, critical questions remain about the nature and impact of effort cues on authenticity perceptions in the context of AI creativity. From an economic perspective, the \textit{authenticity premium} is well documented: market studies demonstrate that, even after adjusting for technical and aesthetic factors, consumers assign higher monetary value to human-made art \cite{Newman2012-yc}. This suggests that economic valuation is deeply tied to perceived human labor, posing strategic challenges for creative industries increasingly reliant on AI tools. Yet most prior investigations have examined human versus AI creativity in broad comparative terms, without systematically isolating which specific markers of effort most strongly influence authenticity judgments \cite{Jordanous2012-mf}. This lack of granularity represents a significant research gap, as understanding the role of visible labor, imperfections, and process narrative could inform both the design of AI creative systems and strategies for integrating them into cultural economies. Addressing this gap is essential not only for refining theoretical models of the effort heuristic in digital creativity, but also for guiding ethical and sustainable adoption of AI in fields where human struggle itself is a key component of perceived value.

To address these theoretical and practical gaps, the present study examines how different forms of process documentation shape perceptions of \textit{authenticity}, \textit{emotional depth}, and \textit{originality} in creative works. Two primary research questions guide this inquiry. \textbf{RQ1} asks:\\
\textit{How do perceptions of authenticity, emotional depth, and originality in creative works change depending on the type of process documentation, and which apparent effort indicators most strongly affect these judgments?}\\
\textbf{RQ2} investigates: \textit{To what extent can process transparency mitigate authenticity disparities between human-authored and AI-generated creative works, and how do evaluations of visible effort cues differ across these domains?}\\
These questions focus on the hierarchical influence of specific effort indicators, such as, process videos, time-tracking records, written explanations, visible imperfections, and work-in-progress materials on authenticity appraisals. Understanding these relationships is critical not only for refining theoretical models of the \textit{effort heuristic} in digital contexts, but also for informing practical strategies in the presentation and marketing of creative works, especially in environments where human and AI contributions coexist.

The primary goal of this study is to examine how visible effort cues shape perceptions of authenticity and value in human- versus AI-generated creative works. Specifically, we investigate (1) which effort indicators most strongly influence authenticity and value attribution, (2) how these cues differentially affect evaluations of human-authored and AI-generated content, and (3) whether effort perception translates into economic premiums for human-made works under varying levels of process transparency. By addressing these questions, the study extends \textit{effort heuristic} theory to contemporary digital creativity and provides empirical insight into how audiences form authenticity judgments in AI-assisted creative contexts. These findings inform both theory and practice, offering guidance for creators, platforms, and curators navigating evolving notions of value in human–AI creative ecosystems.

\section{Methodology}
Motivated by prior work on the \textit{effort heuristic} and authenticity in creative evaluation, we conducted a quantitative cross-sectional survey to examine how visible effort cues shape perceptions of value in human- and AI-generated creative works. This approach allowed us to capture how audiences recognize effort, infer authenticity, and translate these judgments into economic preferences, directly addressing our research questions on process visibility and creative valuation.

\subsection{Research Design}
We employed a quantitative cross-sectional survey design to examine relationships between visible effort cues, perceived authenticity, and economic valuation in human- versus AI-generated creative works. This design enabled the systematic measurement of effort cue recognition, authenticity perceptions, and willingness-to-pay indicators at a single temporal point, supporting statistical analysis of perceptual phenomena \cite{creswell2003research}. The approach aligned with the study’s objective of identifying patterns in effort perception and authenticity judgments across authorship conditions, while ensuring consistency with established \textit{effort heuristic} and AI perception research \cite{Bellaiche2023, kruger2004effort}. The survey questionnaire is available in Section~\ref{sec:SurveyQuestionnaire}.

\subsection{Participants and Sampling Strategy}
Purposive criterion-based sampling was used to recruit 70 participants with established familiarity with digital creative content, following guidelines for focused behavioral phenomena research \cite{creswell2016qualitative, Palinkas2015-uq}. This approach facilitated the selection of information-rich cases directly relevant to the phenomenon of interest and is commonly applied in implementation research. The sample included 51 males (72.9\%) and 19 females (27.1\%), predominantly aged 25–34 years (54.3\%), with most enrolled in bachelor's degree programs (75.7\%). Students constituted the majority (71.4\%), while technical and education professionals each accounted for 10\% of participants. Eligibility required regular engagement with digital creative content, daily or weekly consumption, ensuring adequate familiarity with both human- and AI-generated works, a prerequisite for meaningful evaluation of the study’s creative stimuli and authenticity perception measures.

\subsection{Instruments and Measures}
We designed a structured questionnaire divided into three sections, each targeting constructs central to effort perception and authenticity evaluation.  
\subsubsection{Demographics and Creative Engagement}  
In the first section, we recorded participant characteristics, including age, gender, location, educational background, and occupation. We also collected behavioral data on engagement with creative content, measured on a 5-point scale from daily to rarely, along with purchase history of creative works (4-point scale from regularly to never) and experience in producing creative works, whether professionally, as a hobby, or not at all.  

\subsubsection{Perceptions of Human Effort in Art}  
The second section was developed to assess recognition and evaluation of effort indicators in creative works. We included items measuring the perceived influence of an artist’s personal effort on appreciation (5-point scale from \textit{not at all} to \textit{extremely}); recognition of specific visible effort elements through a multiple-response item listing seven options (process videos, early drawings or notes, mistakes or changes, time spent, tools and materials used, written explanations, live demonstrations); and perceived importance of imperfection in human-authored works (5-point scale from \textit{not at all important} to \textit{absolutely essential}). We also examined the association between visible effort and emotional sincerity (5-point Likert scale) and the extent to which participants believed process evidence should be considered part of the final creative product (5-point scale from \textit{always} to \textit{never}). 

\subsubsection{AI versus Human-Made Creative Work}  
In the final section, we investigated familiarity with and perceptions of AI-generated creative content. We measured familiarity with AI creative tools (4-point scale from \textit{use them} to \textit{not at all}), experience with AI-generated works (yes/no/not sure), and agreement with statements reflecting authenticity concerns: \textit{AI-generated art feels like it’s missing emotional depth or personal touch}, \textit{A perfect output without process feels less meaningful}, and \textit{AI makes it harder to value originality} (all using 5-point Likert scales, adapted from established creative perception measures). We also asked participants to indicate their preference between human works with visible effort and AI works without process (5-point scale from \textit{definitely human} to \textit{definitely AI}) and to report whether they would be willing to pay more for human-made works (binary yes/no).

\subsection{Procedure and Data Analysis}
We briefed all participants on the study objectives and obtained informed consent prior to data collection. Surveys were administered through a combination of online and in-person formats to maximize reach while ensuring a standardized participant experience. Participants evaluated samples of both human- and AI-generated artworks, with stimuli accompanied by visible effort cues designed to prime perceptions before completing structured questionnaires, consistent with established AI-aesthetics research methodologies \cite{Van_Hees2024-os}. Data analysis was conducted using IBM SPSS Statistics version 25. We employed descriptive statistics, including frequencies, percentages, means, and standard deviations, to summarize demographic characteristics and describe central tendencies and variability for key variables such as visible effort, perceived authenticity, and willingness to pay.

\begin{figure}[t]
    \centering
    \includegraphics[width= \textwidth]{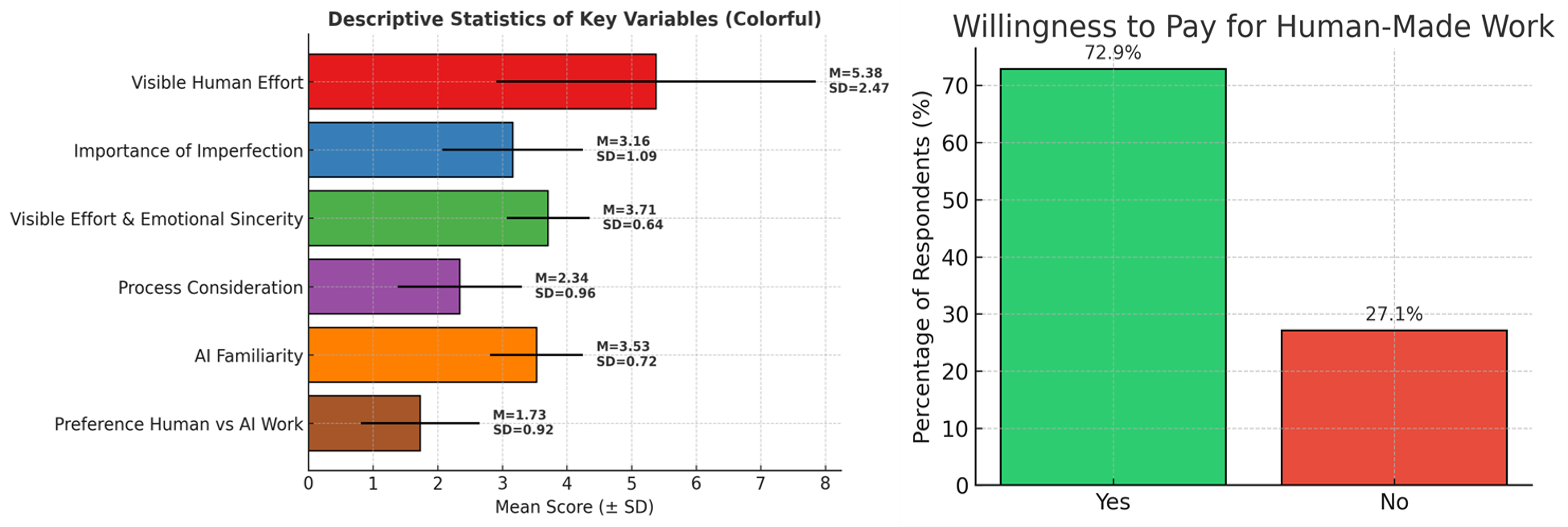}
    \vspace{-2mm}
    \caption{Descriptive Statistics of (a) Key Variables (Mean ± SD) and (b) Willingness to Pay for Human-Made Work vs AI Work}
    \label{fig:descriptive-stats}
    \vspace{-4mm}
\end{figure}

\subsection{Ethical Considerations}
We adhered to established ethical research standards throughout the study, implementing comprehensive protections for participants. Informed consent was obtained from all participants after providing detailed information on study procedures and objectives. Participant anonymity was maintained by collecting responses anonymously and retaining no identifying information within the dataset. Data confidentiality was ensured through secure storage, with access restricted exclusively to research team members. Participants were informed of their right to withdraw from the study at any time without penalty. The study posed no risks to participants’ physical, mental, or social well-being beyond those encountered in everyday life.

\subsection{Participant Demographics}
The study included a total of (N = 70) participants (Table \ref{tab:Demographic}). The sample predominantly consisted of younger adults aged 25–34 years (54.3\%) and males (72.9\%). Most participants held at least a bachelor’s degree (75.7\%) and primarily identified as students (71.4\%), with smaller proportions employed in technical and education-related sectors. Overall, the demographic profile reflects a relatively young, highly educated cohort with substantial exposure to digital and creative environments, making it well aligned with the study’s focus on perceptions of AI-mediated creative content.
The sample size of 70 participants is appropriate for the exploratory and theory-driven goals of a CHI poster contribution, which prioritizes identifying perceptual trends, validating emerging constructs, and probing human–AI interaction phenomena rather than population-level generalization. 

\section{Results}
\subsection{Descriptive Statistics of Key Variables}
Figure \ref{fig:descriptive-stats}a presents the means and standard deviations of key variables related to effort perception and authenticity judgments. Visible human effort received the highest average rating (M=5.38, SD=2.47), indicating that participants strongly recognized and valued visible process cues in creative works. In contrast, preference between human- versus AI-generated work yielded the lowest mean score (M=1.73, SD=0.92), reflecting a relatively low inclination to distinguish based solely on perceived authorship or process visibility. Other constructs such as perceived importance of imperfection (M=3.16, SD=1.09) and the link between visible effort and emotional sincerity (M=3.71, SD=0.64) showed moderate endorsement with less variability, suggesting a more consistent but less polarized viewpoint on these aspects.

\subsection{Perceptions of AI-Generated Creative Works and Visible Effort Elements}
\begin{figure}[t]
    \centering
    \includegraphics[width= \textwidth]{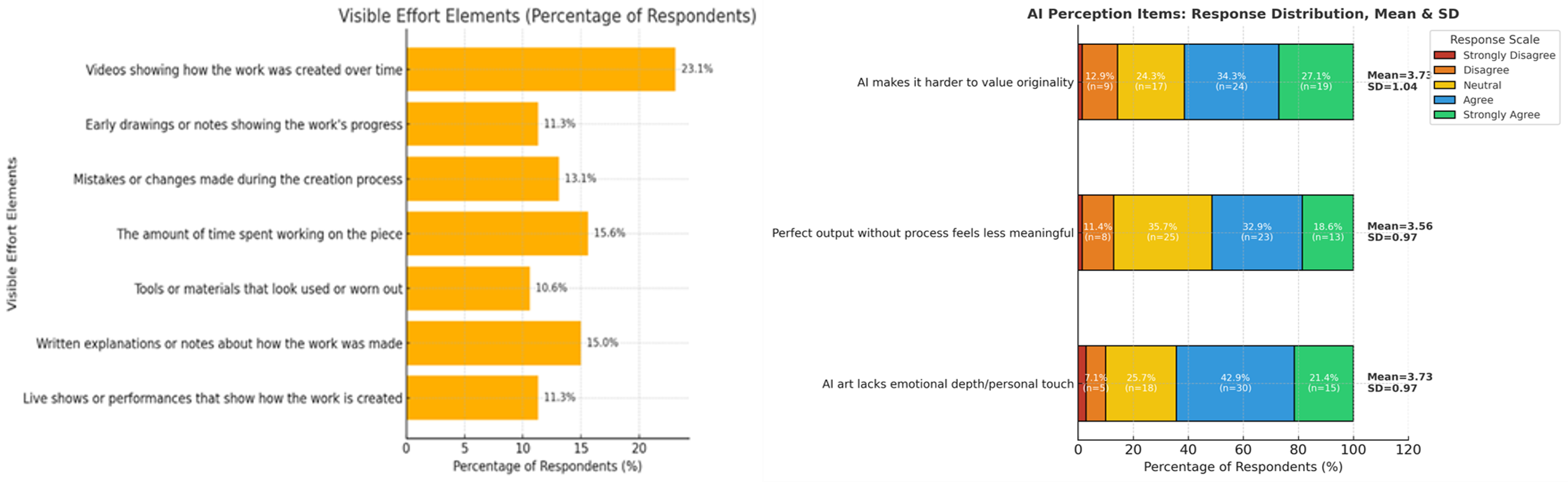}
        \vspace{-2mm}
    \caption{AI Perception Measures: (a) Response Distributions (Mean and Standard Deviation); (b) Recognition Rates of Visible Effort Elements (Percentage of Respondents)}
    \label{fig:fig3-4}
        \vspace{-4mm}
\end{figure}

Figure \ref{fig:fig3-4}a illustrates participant responses to statements regarding AI-generated art and its perceived authenticity. A majority agreed that AI art lacks emotional depth or personal touch (64.3\% combined \textit{agree} and \textit{strongly agree}), and that perfect outputs without visible process feel less meaningful (51.5\% agreement). Additionally, 61.4\% of participants endorsed the view that AI makes it harder to value originality. These findings highlight prevalent skepticism about AI’s capacity to replicate the emotional and original qualities traditionally associated with human creativity.\\
Figure \ref{fig:fig3-4}b presents the frequency of recognition for various visible effort indicators. Videos documenting the creative process were the most commonly identified cue (23.1\%), followed by time spent on the work (15.6\%) and written explanations or notes (15.0\%). Less frequently recognized elements included mistakes or changes made during creation, early drawings or notes, and live demonstrations. This pattern suggests that participants prioritize concrete and traceable evidence of creative labor over more subtle or indirect indicators when attributing effort and authenticity.

\subsection{Willingness to Pay for Human-Made Creative Work}
As shown in Figure \ref{fig:descriptive-stats}b, a substantial majority (72.9\%) of respondents indicated a willingness to pay more for human-made creative works compared to AI-generated counterparts. This result evidences a significant economic \textit{Struggle Premium} associated with perceived human effort and authorship, reinforcing the notion that authenticity judgments carry tangible market implications. The remaining 27.1\% were unwilling to pay a premium, indicating variability in economic valuation potentially linked to individual differences in attitudes toward AI creativity.

\section{Analysis and Discussion}
\subsection{Key Findings}
This study provides empirical evidence supporting the extension of effort heuristic theory \cite{kruger2004effort} to the domain of AI-generated creativity. The findings indicate that visible effort cues, specifically process videos (23.1\% recognition rate), time documentation (15.6\%), and written explanations (15.0\%), have a statistically significant effect on perceptions of authenticity. This outcome aligns with the theoretical prediction that audiences utilize effort as a cognitive shortcut for assessing quality, including in algorithmic contexts.
This extension corresponds with recent theoretical developments in AI perception research. Bellaiche et al. \cite{Bellaiche2023} and Neef et al. \cite{Neef2025-ls} demonstrate that human-created art is systematically \textit{upvalued} rather than AI-generated art being \textit{devalued}, indicating a positive bias toward human creators rather than a general aversion to algorithmic production. Our results further support this \textit{struggle premium} effect, with 72.9\% of participants expressing willingness to pay a premium for human-made work, consistent with the theoretical framework that perceived effort influences value attribution beyond objective quality indicators.
However, the findings also reveal theoretical nuances. Contrary to Newman’s (2012) \cite{Newman2012-yc} emphasis on imperfections as salient markers of authenticity, the data show that mistakes and imperfections were relatively less influential effort cues (ranking lower in Figure \ref{fig:descriptive-stats}b). This suggests a theoretical refinement wherein, within digital creative contexts, audiences may prioritize transparency of the creative process over outcome imperfections when evaluating authenticity. This interpretation aligns with Yusa et al.’s \cite{yusa2022reflections} observation that highlighting human labor during AI-assisted art creation has a stronger impact on perceived authenticity than traditional imperfection-based markers.

\subsection{Addressing Research Questions}
\subsubsection{Research Question 1: Visible Effort Cues and Perceived Authenticity}
Our analysis reveals a clear hierarchy of visible effort cues influencing perceptions of authenticity. Process videos demonstrated the strongest effect (23.1\% recognition rate), followed by time documentation (15.6\%) and written explanations (15.0\%). This distribution indicates that audiences place greater value on procedural transparency, concrete, observable evidence of the creative process, rather than on more abstract or indirect indicators. These findings extend Chamberlain et al.’s (2018) work on emotional resonance by showing that specific, documentable elements of the creative process exert greater influence on authenticity judgments than subjective assessments of quality. Notably, mistakes and imperfections exhibited relatively lower impact, challenging traditional authenticity frameworks that emphasize human fallibility as a central value driver \cite{Newman2012-yc}. Instead, the data suggest that thorough and competent documentation of effort serves as a more salient authenticity cue than indicators of struggle or error. Theoretical implications are substantial: effort heuristics in digital creativity contexts may function differently compared to traditional craft domains. As AI-generated artworks approach \textit{near indistinguishability from human-made works} \cite{Van_Hees2024-os}, audience evaluations appear to prioritize transparency of the creative process over outcome-based quality signals. This shift points to procedural authenticity increasingly supplanting essential authenticity in the evaluation of digital creative outputs.

\subsubsection{Research Question 2: Comparative Evaluations of Human vs. AI Works}
Consistent preference for human-authored works, evidenced by 72.9\% of participants willing to pay a premium, confirms the persistence of human authorship bias across creative domains. However, the findings reveal critical nuances regarding how visible effort cues function differently for human versus AI-generated works. This aligns with emerging research on creativity gatekeeping, which positions human evaluators as pivotal in mediating acceptance of AI-produced artifacts \cite{Magni2024-bi}. The observed moderate increase in positive evaluations of AI-generated works when effort cues were present suggests that process transparency partially mitigates authorship bias but does not fully eliminate the premium associated with human creators. This differential effect carries significant theoretical implications. Notion of \textit{semi-aura} in AI-generated art \cite{Salas_Espasa2025-en} posits that AI works may attain partial authenticity through disclosure of creative process, yet our data indicate this remains subordinate to the value attributed to human authorship. In this context, the effort heuristic operates as a compensatory mechanism for AI-generated works rather than an equivalent pathway to full authenticity.

\subsection{Limitations and Future Directions}
\subsubsection{Theoretical Reflections}
Although the results broadly support effort heuristic theory, several tensions warrant refinement. The relatively low explicit preference for human over AI works (M = 1.73, SD = 0.92) contrasts with stronger willingness-to-pay effects, suggesting that stated preferences and economic valuation may rely on distinct cognitive processes. Additionally, moderate authenticity ratings point to a potential ceiling effect, where visible effort cues yield diminishing returns, implying interaction with other contextual or interpretive factors.

\subsubsection{Implications and Future Work}
Practically, the strong influence of process videos and time documentation highlights the value of process transparency in establishing authenticity for both human and AI-assisted creators. This aligns with broader calls for AI transparency and explainability as mechanisms for trust-building \cite{ibm2025}. While prior work shows that AI labeling can negatively affect credibility and creativity judgments \cite{Cheong2025-ud}, our findings suggest that rich process disclosure may partially offset these effects.
Future work should examine more diverse populations \cite{Balasubramaniam2023-jf} and move beyond stated preferences toward behavioral and longitudinal methods, including real economic transactions, to better capture how effort perception shapes value in real-world creative contexts.

\section{Conclusion}
This study examines the \textit{Struggle Premium} and finds that visible effort cues, especially process transparency, play a meaningful role in how audiences judge authenticity and value in AI-era creativity. While human-authored works retain a strong premium, transparent process disclosure can partially mitigate the devaluation of AI-generated content, pointing to new forms of creative authenticity. Extending effort heuristic theory, our findings suggest a shift from outcome-based authenticity toward procedural transparency in digital creativity. The coexistence of a persistent human premium with only moderate gains for AI art indicates that authenticity is multi-dimensional rather than a simple human–AI continuum. These insights inform creative practice, platform design, and cultural policy in contexts of growing human–AI collaboration.

\begin{acks}
We express our sincere gratitude to \href{https://ciol-researchlab.github.io/}{Computational Intelligence and Operations Laboratory (CIOL)} for their invaluable guidance, unwavering support, and continuous assistance throughout this journey.
\end{acks}

\bibliographystyle{ACM-Reference-Format}
\bibliography{arxiv}

\appendix

\section{Survey Questionnaire} \label{sec:SurveyQuestionnaire}

\noindent\textbf{Note:} Bengali translations (not shown here) were included in the original questionnaire for reader comprehension.

\subsection{Consent for Participation and Data Use}
Consent statement regarding purpose of data collection, anonymity, voluntary participation, and confidentiality.
By selecting I ACCEPT, participants confirm they have read and understood the consent statement and agree to participate.

$\boxdot$ I ACCEPT

\subsection{Demographic Data}

\begin{enumerate}
    \item Age  
    \begin{itemize}
        \item Under 18
        \item 18--24
        \item 25--34
        \item 35--44
        \item 45--54
        \item 55--64
        \item 65+
        \item Other
    \end{itemize}
    \item Gender  
    \begin{itemize}
        \item Male
        \item Female
        \item Prefer not to say
        \item Other
    \end{itemize}
    \item Location (Country/Region)  
    \begin{itemize}
        \item Bangladesh
        \item Other
    \end{itemize}
    \item Educational Background  
    \begin{itemize}
        \item No formal education
        \item High school
        \item College
        \item Bachelor's degree
        \item Postgraduate degree
        \item Other
    \end{itemize}
    \item Occupation/Field of Work  
    \begin{itemize}
        \item Student
        \item Creative professional (e.g., designer, artist, writer)
        \item Technical professional (e.g., engineer, IT)
        \item Business/Finance
        \item Education (e.g., teacher)
        \item Healthcare (e.g., doctor, nurse, social worker)
        \item Retired
        \item Other
    \end{itemize}
\end{enumerate}

\subsection{Questions}

\begin{enumerate}
    \item How often do you engage with creative content (art, music, writing, etc.)?
    \begin{itemize}
        \item Daily
        \item A few times a week
        \item Weekly
        \item Monthly
    \end{itemize}
    
    \item Have you ever purchased creative work (art, crafts, music, writing, etc.)?
    \begin{itemize}
        \item Yes, regularly
        \item Occasionally
        \item Rarely
        \item Never
    \end{itemize}
    
    \item Do you yourself create any art or creative work?
    \begin{itemize}
        \item Yes, professionally
        \item Yes, as a hobby
        \item No
    \end{itemize}
    
    \item When evaluating creative work, how much does the artist’s personal effort influence your appreciation?
    \begin{itemize}
        \item Not at all
        \item Slightly
        \item Moderately
        \item Significantly
        \item Extremely
    \end{itemize}
    
    \item Which elements make effort most visible to you? (Select all that apply)
    \begin{itemize}
        \item Videos showing how the work was created over time
        \item Early drawings or notes showing the work's progress
        \item The artist’s personal story or feelings that helped create the work
        \item Mistakes or changes made during the creation process
        \item The amount of time spent working on the piece
        \item Tools or materials that look used or worn out
        \item Written explanations or notes about how the work was made
        \item Live shows or performances that show how the work is created
        \item Other
    \end{itemize}
    
    \item How important is the concept of imperfection in your definition of human-authored art?
    \begin{itemize}
        \item Not at all important
        \item Slightly important
        \item Moderately important
        \item Very important
        \item Absolutely essential
    \end{itemize}
    
    \item Do you associate visible effort with emotional sincerity?
    \begin{itemize}
        \item Strongly agree
        \item Agree
        \item Neutral
        \item Disagree
        \item Strongly disagree
    \end{itemize}
    
    \item Would you consider process/evidence of effort to be part of the final ``product'' of creative work?
    \begin{itemize}
        \item Always
        \item Often
        \item Sometimes
        \item Rarely
        \item Never
    \end{itemize}
    
    \item Are you familiar with AI tools for generating art, music, or writing (e.g., Midjourney, ChatGPT, Suno)?
    \begin{itemize}
        \item Yes, I use them
        \item Yes, I’ve seen them used
        \item I’ve heard of them
        \item No, not at all
    \end{itemize}
    
    \item Have you ever bought, downloaded, or appreciated a piece of AI-generated creative work?
    \begin{itemize}
        \item Yes
        \item No
        \item Not sure
    \end{itemize}
    
    \item To what extent do you agree with the following statements? (Likert scale: 1 = Strongly Disagree, 5 = Strongly Agree)
    \begin{itemize}
        \item I find AI-generated art visually impressive
        \item AI-generated art feels like it’s missing emotional depth or a personal touch
        \item I think it matters if a piece is created by a human or AI
        \item I feel connected to the process behind human-made work
        \item A perfect output without process feels less meaningful
        \item AI makes it harder to value originality
    \end{itemize}
    
    \item If you had to choose between two similar works (one by a human with visible effort, and one by AI with no process), which would you prefer?
    \begin{itemize}
        \item Definitely human
        \item Probably human
        \item No preference
        \item Probably AI
        \item Definitely AI
    \end{itemize}
    
    \item Would you be willing to pay more for the human-made work in that scenario?
    \begin{itemize}
        \item No
        \item Yes
    \end{itemize}
    
    \item What does authenticity in creative work mean to you?
    
    \item Can AI-generated work ever feel emotionally real to you? Why or why not?
    
    \item Do you think AI creators should highlight their effort and process more explicitly in the future? How might this affect your experience as an audience or buyer?
    
    \item Any additional comments or thoughts about human effort vs. AI in creativity?
\end{enumerate}

\section{Survey Data and Results}

\begin{table}[htbp]
\centering
\caption{Demographic Profile of the Respondents}
\label{tab:Demographic}
\begin{tabular}{|l|l|c|c|}
\hline
\textbf{Variable} & \textbf{Category} & \textbf{Freq (n)} & \textbf{Percent (\%)} \\
\hline
\multirow{5}{*}{\shortstack[l]{Age}}
& 18--24 & 28 & 40.0 \\
& 25--34 & 38 & 54.3 \\
& 35--44 & 1  & 1.4  \\
& 45--54 & 2  & 2.9  \\
& 55--64 & 1  & 1.4  \\
\hline
\multirow{2}{*}{Gender}
& Male   & 51 & 72.9 \\
& Female & 19 & 27.1 \\
\hline
\multirow{2}{*}{Education}
& Bachelor’s   & 53 & 75.7 \\
& Postgraduate & 17 & 24.3 \\
\hline
\multirow{6}{*}{Occupation}
& Student               & 50 & 71.4 \\
& Creative Professional & 2  & 2.9  \\
& Technical Professional& 7  & 10.0 \\
& Business/Finance      & 3  & 4.3  \\
& Education             & 7  & 10.0 \\
& Healthcare            & 1  & 1.4  \\
\hline
\multirow{4}{*}{\shortstack[l]{Frequency of Creative\\Content Engagement}}
& Daily              & 21 & 30.0 \\
& A few times a week & 26 & 37.1 \\
& Weekly             & 7  & 10.0 \\
& Monthly            & 16 & 22.9 \\
\hline
\multirow{4}{*}{\shortstack[l]{Creative Work \\Production or \\Purchase History}}
& Regularly    & 4  & 5.7  \\
& Occasionally & 30 & 42.9 \\
& Rarely       & 22 & 31.4 \\
& Never        & 14 & 20.0 \\
\hline
\multirow{3}{*}{Personal Creative Work Creation}
& Professionally & 4  & 5.7  \\
& As a hobby     & 34 & 48.6 \\
& No             & 32 & 45.7 \\
\hline
\end{tabular}
\end{table}

\end{document}